\documentclass[footinbib,aps,pra,reprint,superscriptaddress,nopacs]{revtex4-1}

\usepackage{amsmath,amssymb,upgreek,bm,verbatim}
\usepackage[absolute]{textpos}
\usepackage{graphicx, tabularx, braket}
\usepackage{xcolor}

\usepackage[absolute]{textpos}
\usepackage{multirow}

\definecolor{darkblue}{rgb}{0,0,0.5}

\usepackage{hyperref}
\hypersetup{
colorlinks=true,
linkcolor=darkblue,
filecolor=blue,
citecolor=darkblue,  
urlcolor=darkblue,
}

\newcommand\bl[1]{\textcolor{black}{#1}}

\begin{document}
\title{On-chip frequency-bin quantum photonics}

\author{Karthik V. Myilswamy}
%\thanks{These authors contributed equally to this work.}
\affiliation{School of Electrical and Computer Engineering and Purdue Quantum Science and Engineering Institute, Purdue University, West Lafayette, Indiana 47907, USA}
\affiliation{National Institute of Standards and Technology, Boulder, Colorado 80305, USA}

\author{Lucas M. Cohen}
%\thanks{These authors contributed equally to this work.}
\affiliation{School of Electrical and Computer Engineering and Purdue Quantum Science and Engineering Institute, Purdue University, West Lafayette, Indiana 47907, USA}
\affiliation{Photonic and Phononic Microsystems, Sandia National Laboratories, Albuquerque, New Mexico 87123, USA}

\author{Suparna Seshadri}
%\thanks{These authors contributed equally to this work.}
\affiliation{School of Electrical and Computer Engineering and Purdue Quantum Science and Engineering Institute, Purdue University, West Lafayette, Indiana 47907, USA}
\affiliation{Aliro Technologies, Inc., Brighton, Massachusetts 02135, USA}

\author{Hsuan-Hao Lu}
\affiliation{Quantum Information Science Section, Computational Sciences and Engineering Division, Oak Ridge National Laboratory, Oak Ridge, Tennessee 37831, USA}

\author{Joseph M. Lukens}
\affiliation{School of Electrical and Computer Engineering and Purdue Quantum Science and Engineering Institute, Purdue University, West Lafayette, Indiana 47907, USA}
\affiliation{Quantum Information Science Section, Computational Sciences and Engineering Division, Oak Ridge National Laboratory, Oak Ridge, Tennessee 37831, USA}
\affiliation{Research Technology Office and Quantum Collaborative, Arizona State University, Tempe, Arizona 85287, USA}

\date{\today}

\begin{abstract}
Frequency-bin encoding furnishes a compelling pathway for quantum information processing systems compatible with established lightwave infrastructures based on fiber-optic transmission and wavelength-division multiplexing. Yet although significant progress has been realized in proof-of-principle tabletop demonstrations, ranging from arbitrary single-qubit gates to controllable multiphoton interference, challenges in scaling frequency-bin processors to larger systems remain. In this Perspective, we highlight recent advances at the intersection of frequency-bin encoding and integrated photonics that are fundamentally transforming the outlook for scalable frequency-based quantum information. Focusing specifically on results on sources, state manipulation, and hyperentanglement, we envision a possible future in which on-chip frequency-bin circuits fulfill critical roles in quantum information processing\bl{, particularly in communications and networking.}
\end{abstract}

\maketitle

\section{Introduction}
\label{sec:intro}
One of several reasons for the remarkable success of lightwave communications,
wavelength-division multiplexing (WDM) has continuously asserted its importance in classical optical networks~\cite{Agrell2024}: first in coarse WDM (CWDM) based on 20~nm-wide channels~\cite{ITU2003}, followed by dense WDM (DWDM) with spacings as tight as 12.5~GHz~\cite{ITU2020}, and finally evolving into the current flex-grid era of programmable frequency slots~\cite{Gerstel2012, Jinno2017}. The optically transparent nature of WDM technology---i.e., its avoidance of optical-to-electrical conversion---has likewise solidified its place in the quantum domain, with uses including quantum--classical coexistence~\cite{Townsend1997b,  Peters2009, Chapuran2009, Eraerds2010, Patel2012} and broadband entanglement distribution~\cite{Lim2008, Herbauts2013, Wengerowsky2018, Lingaraju2021}.

Yet additional features of the frequency degree freedom (DoF)---such as its ease of entanglement generation, stability in long-haul fiber transmission, and compatibility with stationary qubits based on distinct energy levels~\cite{Moehring2007, Lan2007}---have inspired efforts to expand its role from \emph{multiplexing} quantum information alone to \emph{encoding} quantum information itself~\cite{Lukens2017, Kues2019, Lu2019c, Lu2023c}. In the discrete-variable version of this paradigm, single photons occupying superpositions of $d$ discrete frequency modes, or bins, form quantum-information-carrying ``qudits,'' with qubits the special case $d=2$.

Although quantum operations for such qudits is decidedly nontrivial, requiring controllable multifrequency interference, the last decade has witnessed remarkable experimental progress in frequency-bin processing, including nonlinear-~\cite{Kobayashi2016, Clemmen2016, Kobayashi2017} and electro-optic~\cite{Lu2018a, Imany2018c, Lu2018b, Kashi2021, Xue2022} frequency beamsplitters, full state characterization enabled by spectro-temporal interference~\cite{Kues2017, Imany2018a, Lu2022b, Clementi2023, Borghi2023}, and unitary quantum gates via alternating electro-optic phase modulators (EOMs) and Fourier-transform pulse shapers~\cite{Lu2018a, Lu2018b, Lu2019a, Lu2020b, Lu2022a, Lingaraju2022, Lu2023a, Henry2024}. Indeed, the last approach---known as the quantum frequency processor (QFP)---has been shown theoretically capable of scalable universal quantum information processing~\cite{Lukens2017}, thus bequeathing the field of frequency-bin quantum information with a framework for general-purpose quantum computation\bl{, above and beyond its core synergies in the subfield of quantum communications and networking.}

Despite the variety and success of proof-of-principle demonstrations, however, progress toward the ultimate vision of frequency-bin processing is reaching bounds imposed by tabletop components, whose insertion loss and cost have thus far limited experimental QFPs to three elements (EOM/pulse shaper/EOM) \bl{and whose size and limited bandwidth (both optical and microwave) have hampered scaling to many photons and higher dimensions.}  Of course, the constraints of tabletop optics are nothing new; both the general issues of scalability and the ultimate solution of photonic integration are shared by more traditional optical qubits, with path encoding arguably the most developed~\cite{Politi2008, Wang2018a, Wang2020} due to straightforward gate construction with beamsplitters and phase shifters~\cite{Reck1994, Clements2016}. \bl{Interested readers may refer to other review articles summarizing quantum information processing in other DoFs~\cite{erhard2020, xavier2020, nape2023} and the transformative role now played by integrated photonics~\cite{Chen2021b, Moody2022, labonte2024, kaur2024}.} 

%Although this perspective is primarily focused on frequency-bins, there are other DOFs including polarization, time-bins, path, and spatial modes which are also commonly explored for quantum encoding. 

Notwithstanding its comparative immaturity, frequency bins do enjoy unique features that make them ideal candidates for integrated photonics. At a high level, their ability to encode and multiplex high-dimensional qudits in a single waveguide could provide a pathway to even more compact footprints than path-encoded alternatives~\cite{Kues2019}. And practically speaking, frequency bins are arguably the most natural encoding format for microring resonators---omnipresent components in silicon (Si) photonics~\cite{Bogaerts2012}. Not only do microring filters provide the functionalities needed for on-chip multiplexing~\cite{Dahlem2011, Cohen2024a} and pulse shaping~\cite{Agarwal2006, Khan2010, Wang2015b, Nussbaum2022, Cohen2024b}, but they also automatically produce frequency-bin-entangled photons in resonantly enhanced spontaneous four-wave mixing (SFWM)~\cite{Clemmen2009, Chen2011, Azzini2012}.

In this Perspective, we overview exciting developments in integrated quantum photonics that engender optimism for a future of scalable frequency-bin quantum information processing. After first summarizing frequency-bin-entangled sources demonstrated in a variety of material platforms in Sec.~\ref{sec:generation}, we transition to on-chip manipulation technology in Sec.~\ref{sec:manipulation}. While generally more complex for the frequency DoF compared to spatial encodings, recent results with microring pulse shapers, integrated EOMs, and the photonic molecule have laid a promising foundation for complete frequency processors on chip. Finally, in Sec.~\ref{sec:hyperent} we explore the union of frequency-bin entanglement with other DoFs via hyperentanglement. 

\bl{While we are optimistic about the impact of frequency encoding in quantum networks in particular, where its stability and parallelism in optical fiber form a persuasive combination, the extent of its role (e.g., in multiplexing, signal encoding, or both) is admittedly difficult to predict. Yet regardless of the ultimate home for frequency bins in quantum information science,} this Perspective argues for photonic integration as a critical step along the way: necessary, though not sufficient, to realize its full potential.

\section{State generation} 
\label{sec:generation}

Many early demonstrations of biphoton frequency combs (BFCs) exploited spontaneous parametric downconversion (SPDC) in $\chi^{(2)}$ bulk crystals or waveguides. Unless already generated inside a cavity, these biphotons were subsequently shaped into BFCs using post-generation spectral filtering techniques, such as etalons or programmable filters, resulting in the loss of nonresonant photon pairs. 
\bl{In recent years, more intricate techniques with engineered phase matching \cite{Morrison2022, Chiriano2023, Shukhin2024} and time-varying cavities~\cite{myilswamy2024} have also been explored for BFC generation.
Yet in the context of integrated photonics, it is the microring resonator which has emerged as the defining BFC source, uniting a compact footprint with direct frequency-bin generation in resonant modes.} %of the microrings define the frequency bins, allowing photon pairs to be generated at these resonances directly and eliminating the losses associated with post-generation filtering. 
This concept was first theoretically explored in Si-on-insulator (SOI) microrings which, as $\chi^{(3)}$ sources, generate photon pairs through SFWM~\cite{Chen2011}.

\subsection{Single microring}

\begin{figure}[tb!]
\centering\includegraphics[trim= 0 75 0 0,clip,width=3in]{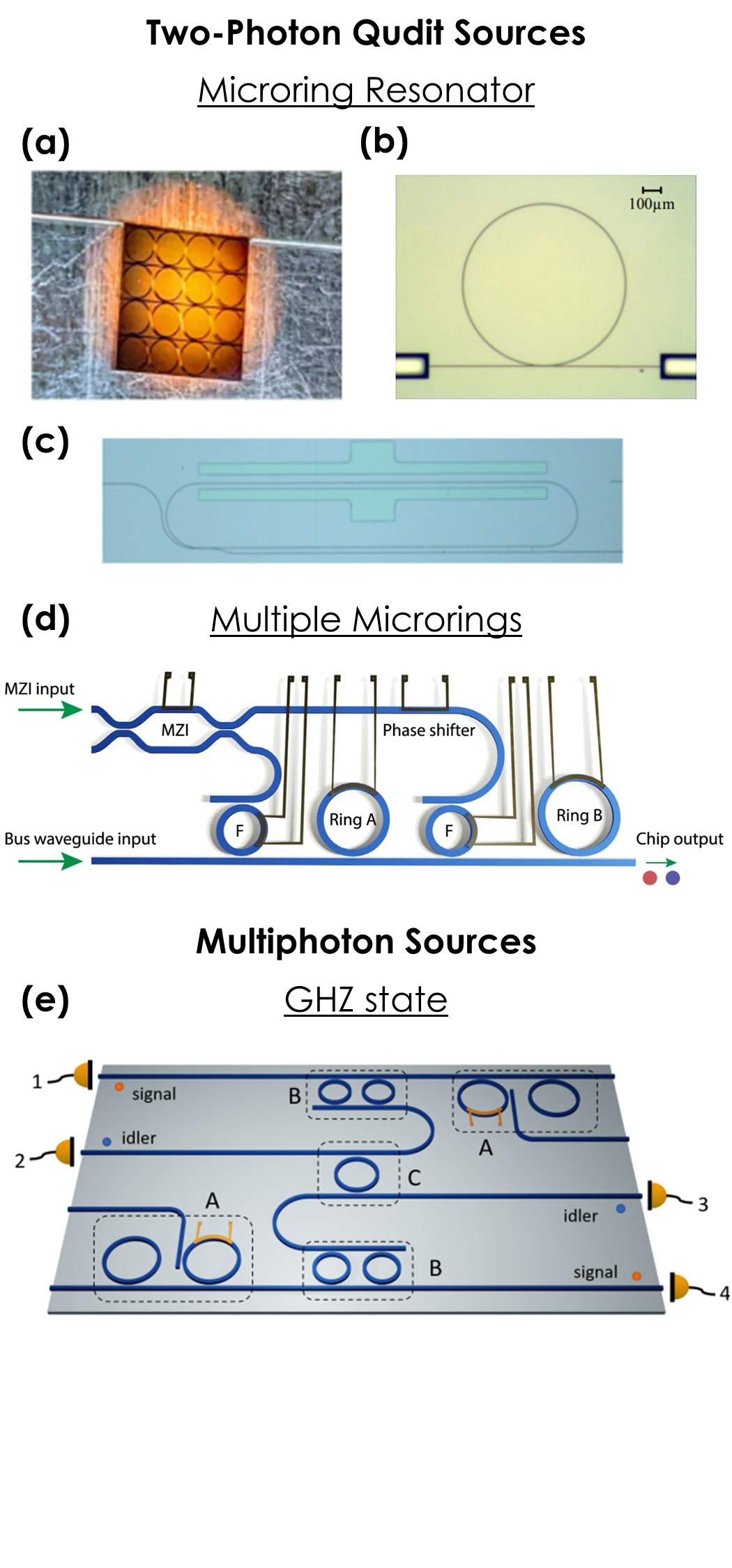}
\caption{Integrated sources of frequency-bin-entangled qudits. SiN microrings with (a)~40~GHz \cite{Lu2022b} and (b)~50~GHz \cite{Imany2018a} FSRs. (c) TFLN source with a 32~GHz FSR~\cite{Seshadri2024}. (d)~Si-based dual-microring source with 377.2 and 373.4 GHz FSRs~\cite{Clementi2023}. (e)~Design for producing a four-photon \bl{Greenberger--Horne--Zeilinger (GHZ)} state~\cite{Banic2024a}. \bl{The working principles of (d,e) are elaborated in the text.} Images reproduced with permission from a Creative Commons Attribution 4.0 International License (\url{http://creativecommons.org/licenses/by/4.0/})~\cite{Lu2022b, Clementi2023}, an Optica Publishing Group Open Access License~\cite{Imany2018a}, and the American Physical Society~\cite{Banic2024a}.}
\label{fig:MRR}
\end{figure}

The simplest integrated geometry for BFC generation involves a single microring coupled to a bus waveguide, with several examples illustrated in Fig.~\ref{fig:MRR}(a--c)~\cite{Imany2018a,Lu2022b,Seshadri2024}. By coupling the ring to a second waveguide as well, photon pairs can be extracted from both the through and drop ports~\cite{Kues2017}. These demonstrations predominantly utilize SFWM in platforms such as Si~\cite{Clemmen2009, Azzini2012, Engin2013, Savanier2016, Alexander2024}, silicon nitride (SiN)~\cite{Jaramillo2017,Imany2018a,Imany2018b,Samara2019,Joshi2020,Wu2021,Lu2022b,Mahmudlu2023,Chen2024}, Hydex~\cite{Reimer2014,Kues2017,Roztocki2017,Reimer2019,Sugiura2020}, and aluminum gallium arsenide (AlGaAs)~\cite{Steiner2021}. 

Although inherent frequency-bin entanglement had long been theorized in such platforms~\cite{Chen2011}, it was not experimentally confirmed until the advent of advanced characterization techniques utilizing electro-optic frequency mixing. Reference~\cite{Kues2017} represents a pioneering effort in this area, generating BFCs with a 200~GHz free spectral range (FSR) through a Hydex microring and fully reconstructing two-photon density matrices within a $4 \times 4$ two-qudit Hilbert space. \bl{However, such large FSRs far exceed the bandwidth of typical commercial EOMs, %which are typically limited to 50 GHz at the time, 
making integration with emerging electro-optic-based quantum frequency gates challenging.}

Continuous efforts have since focused on demonstrating BFCs at smaller FSRs. For instance, SiN microrings with FSRs of 40~GHz [Fig.~\ref{fig:MRR}(a)]~\cite{Lu2022b} and 50~GHz [Fig.~\ref{fig:MRR}(b)]~\cite{Imany2018a}, as well as a spiral Si microring with a 21 GHz FSR~\cite{Henry2024}, have pushed bin spacings into a regime much more compatible with electro-optic manipulation, facilitating measurement and inference of  BFC density matrices up to ($8 \times 8$)-dimensional two-qudit Hilbert spaces, the highest fully characterized frequency-bin dimension to date~\cite{Lu2022b}. However, a fundamental tradeoff exists in single microring systems, where the photon pair generation rate scales as $\frac{Q^3}{R^2}$---with $Q$ representing the quality factor and $R$ the microring radius---indicating that lower FSR rings produce lower pair generation rates at the same pump power~\cite{Chembo2016}. In Sec.~\ref{multipleMRR}, we will explore recent proposals aimed at overcoming this tradeoff.

%In Ref.~\cite{Kues2017}, BFCs were generated using a Hydex microring with a free spectral range (FSR) of 200~GHz, with two-photon density matrices reconstructed in up to a $4\times 4$ two-qudit Hilbert space. However, such large FSRs pose challenges for electro-optic manipulation of qudit states, as commercial electro-optic modulators typically have a bandwidth of only 50~GHz. In contrast, Refs.~\cite{Imany2018a} [Fig.~\ref{fig:MRR}(b)] and~\cite{Lu2022b} [Fig.~\ref{fig:MRR}(a)] demonstrate BFCs using silicon nitride microrings with FSRs of 50 GHz and 40 GHz, respectively, making them much more suitable for electro-optic manipulation. However, the photon pair generation rate scales with $\frac{Q^3}{R^2}$, where $Q$ is the quality factor and $R$ is the microring radius, indicating that lower FSR rings yield lower pair generation rates for the same pump power~\cite{Chembo2016}. Notably, in Ref.~\cite{Lu2022b}, the density matrices of the BFCs were reconstructed up to an $8 \times 8$ two-qudit Hilbert space, the highest frequency-bin dimension reported to date.

While $\chi^{(3)}$ microrings are frequently investigated for BFC generation, the photon pairs and the pump light occupy the same optical band (the 1530--1565~nm telecom C-band in most demonstrations), which complicates the pump suppression necessary for a high coincidences-to-accidentals ratio (CAR). Ideally, a pump rejection of $\gtrsim$100~dB is desired, which is challenging but can be accomplished with dedicated on-chip filter networks~\cite{Afifi2021, Goswami2022,Mahmudlu2023,Alexander2024}. %In silicon sources, p-i-n junctions are employed to mitigate performance degradations related to free carriers~\cite{Engin2013}. Additionally, significant contamination occurs at the wavelengths of the photon pairs due to strong spontaneous Raman scattering of the pump light in optical fibers, which tends to spread across the entire C-Band~\cite{Samara2019} and cannot be easily avoided. One strategy to alleviate this issue is cooling the fibers~\cite{Joshi2022}. 

In contrast, $\chi^{(2)}$ sources effectively mitigate this issue as the photon pairs are generated through SPDC, with the pump typically far away in the 780~nm band. % and the generated photon pairs in the telecom band. 
This dramatically simplifies pump suppression, making $\chi^{(2)}$ sources particularly attractive for achieving higher CAR. $\chi^{(2)}$ microrings have been implemented on thin-film lithium niobate (TFLN)~\cite{Ma2020,Javid2023,Hwang2024,Seshadri2024} and aluminum nitride (AlN)~\cite{Guo2017b,Ma2020,Zhang2023}. However, \bl{given the wide bandwidth separation of the interacting fields, additional efforts are often required to achieve phase matching and simultaneous resonance, such as spatial mode matching~\cite{Zhang2023,Guo2017b}, periodic poling~\cite{Javid2023,Seshadri2024,Hwang2024}, or linearly uncoupled rings~\cite{Zatti2022}}. Figure~\ref{fig:MRR}(c) depicts a periodically poled TFLN microring source with a 32~GHz FSR, where frequency-bin entanglement has been confirmed up to a $3\times3$ state~\cite{Seshadri2024}. %Additionally, TFLN presents a promising platform for frequency qudits due to its exceptional electro-optic properties~\cite{Boes2023}, facilitating easy on-chip mixing of frequency bins essential for state manipulation. , while also allowing for heterogeneous integration with CMOS platforms~\cite{Wang2023a}.

%\fix{References to more recent works: Junqui Liu~\cite{chen2024ultralow}, }
%\fix{ \\
%- Thin-film lithium niobate rings for generating entangled photons: \cite{zhang2023, hwang2024spontaneous, ma2020ultrabright, javid2023chip, seshadri2024measuring} \\
%- heterogeneous integration of LN with CMOS compatible platforms~\cite{chang2017heterogeneous, ahmed2018vertical, churaev2023heterogeneously}}.

\subsection{Multiple microrings}
\label{multipleMRR}
As mentioned earlier, microrings with higher FSRs offer higher pair generation rates and are more compact compared to their lower FSR counterparts. However, FSRs in the range of hundreds of GHz are inefficient in bandwidth utilization and ill-suited for electro-optic manipulation. To address the tradeoff between frequency-bin spacing and pair generation rate\bl{---as well as enhance flexibility in BFC state engineering---}multi-microring approaches have been pursued~\cite{Liscidini2019,Sabattoli2022,Clementi2023,Borghi2023}. In this scenario, multiple microrings with higher FSRs are pumped simultaneously, with each microring contributing one frequency-bin pair, creating a high-dimensional BFC via coherent superposition. The frequency-bin spacing is no longer tied to the ring radius---and thus the pair generation rate---but is instead determined by the FSR difference between microrings. An example device is illustrated in Fig.~\ref{fig:MRR}(d), featuring two Si microrings with FSRs of 377.2 and 373.4~GHz, producing 19~GHz frequency qubits located five FSRs away from the pump~\cite{Clementi2023}. %Each ring produces one frequency-bin pair of a two-dimensional frequency qubit state. 
By tuning the rings for simultaneous pumping at the same resonance frequency and using heaters to adjust the relative phase between the rings, two of the four Bell states with negative correlations were demonstrated. By detuning the rings and pumping them at different wavelengths, the remaining two Bell states with positive correlations were also achieved---providing a complementary, integrated version of complete frequency-bin Bell state synthesizers in tabletop systems~\cite{Seshadri2022}. %This approach not only eliminates the trade-off between generation rate and frequency-bin spacing but also enables programmable generation of specific Bell states. 
%Notably, single microrings typically generate one of the four Bell states with negative correlations and require further state manipulation to transform them into the other Bell states. 
By simply adding more microrings, higher-dimensional qudit states can also be generated~\cite{Borghi2023}. %, a similar approach involving four microrings was implemented, enabling the generation of four-dimensional qudit states. 
Although the number of rings scales with qudit dimension, this approach offers greater programmability and flexibility than single-ring BFCs, along with higher pair generation rates at smaller frequency-bin spacings.

\subsection{Multiphoton state generation}
The previous subsections focus on two-photon frequency-bin entangled states. However, multipartite entangled states, where entanglement extends beyond two photons, hold significant interest for both fundamental studies and practical applications. These states are valuable for distributed quantum computing~\cite{Cirac1999} and multiparty quantum communication protocols~\cite{Bose1998,Hillery1999}. One classic example is the Greenberger--Horne--Zeilinger (GHZ) state~\cite{Greenberger1990}, where all parties are maximally entangled as a whole, but none of the reduced subsystems exhibit entanglement. For example, a four-party GHZ state in two dimensions can be represented as $\ket{0000} + \ket{1111}$. Early demonstrations of GHZ states concentrated on the polarization DoF~\cite{Lu2009, Bouwmeester1999}. In one such approach~\cite{Bouwmeester1999}, two entangled photon pairs are generated, and a fusion operation~\cite{Kok2007} is applied to one photon from each pair. %, followed by postselection of events leading to a four-photon GHZ state. 
For the polarization DoF, this fusion is realized by a polarizing beamsplitter (PBS) followed by postselecting on one photon appearing in each spatial mode. 

%Early demonstrations of GHZ states were in the polarization DoF~\cite{Lu2009,Bouwmeester1999}, and the generation approaches can be broadly divided into two categories.
%In the first method~\cite{Lu2009}, $n$ two-photon entangled pairs are prepared, and a GHZ state projection is performed on $n$ photons, one from each pair, resulting in an $n$-photon GHZ state from the remaining photons. 

An elegant translation of this approach has been proposed, with a microring filter as the frequency-bin analog of the PBS as shown in Fig.~\ref{fig:MRR}(e)~\cite{Banic2024a}. In this design, the two entangled photon pair sources are represented by dashed boxes labeled $A$, each formed by a two-ring system (cf. Sec.~\ref{multipleMRR}). The signal-idler photon pairs are directed to two microrings tuned to the idler bins (dashed box $B$), where the signal and idler photons are spatially demultiplexed. The idler photons from both sources then converge at a central fusion microring (dashed box $C$), which swaps the waveguide for one idler frequency bin while passing the other, effectively mimicking the PBS in the frequency domain. Another important class of multipartite states, known as $W$ states, are robust against photon loss, as their reduced subsystems preserve entanglement~\cite{Dur2000, Bouwmeester1999}. Using spectral filtering and postselection similar to the GHZ approach, Ref.~\cite{Banic2024a} also discusses methods for synthesizing frequency-bin $W$ states on photonic integrated circuits.

\bl{The examples above are limited to three- and four-photon states, with each photon restricted to two-dimensional systems. Reference~\cite{Banic2024b} proposes a theoretical framework to generalize these schemes to higher dimensions and larger photon numbers by combining multiple photon-pair sources with the application of complex unitary frequency-bin operations. Ultimately, the generation rate of these synthesized states depends on the photon-pair rates of individual sources and the success probability of projective measurements---challenges that impact all photonic DoFs. In general, active multiplexing of entangled photon sources~\cite{meyer2020} and the use of solid-state single-photon emitters~\cite{Aharonovich2016}, as alternatives to traditional SFWM and SPDC sources, offer promising approaches to circumvent the inherent tradeoff between pair rates and CAR~\cite{Takesue2010}, thereby enabling more efficient multipartite entangled state generation~\cite{Pont2024}.}

\subsection{High spectral purity sources}
\begin{figure}[tb!]
\centering\includegraphics[trim= 0 160 0 0,clip,width=3in]{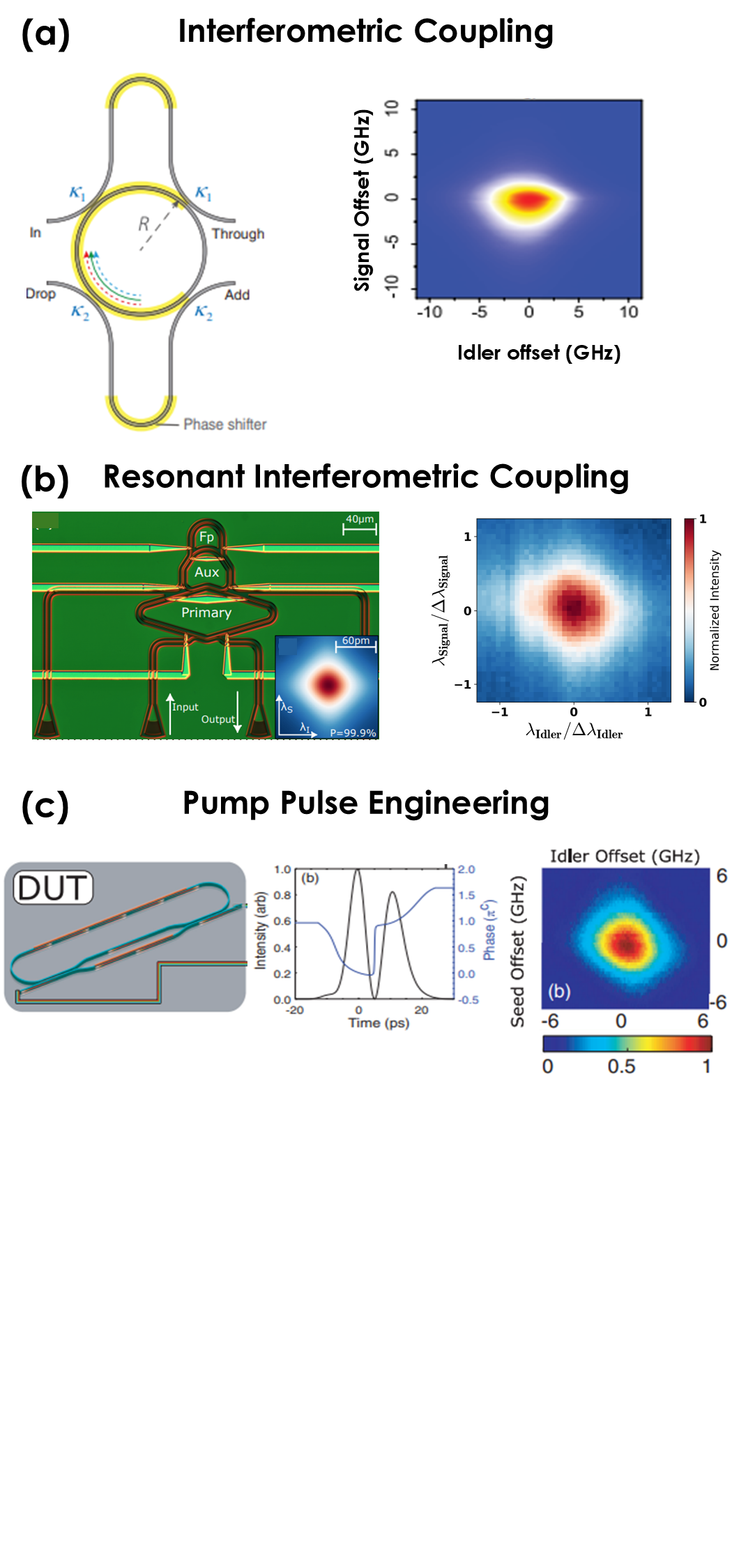}
\caption{Integrated demonstrations of spectrally pure photon pairs. (a)~Interferometric coupling~\cite{Liu2019}. (b) Resonantly controlled interferometric coupling~\cite{Burridge2023}. (c)~Tailored pump pulses~\cite{Burridge2020}. Device designs appear in the left column, with corresponding JSIs in the right column. In (c), the pump pulse is also illustrated in the center column. 
Images reproduced with permission from a Creative Commons Attribution 4.0 International License (\url{http://creativecommons.org/licenses/by/4.0/})~\cite{Burridge2023} and Optica Publishing Group~\cite{Liu2019,Burridge2020}.}
\label{fig:purity}
\end{figure}

Multiphoton interference between independent sources is central to many quantum information protocols, such as quantum teleportation~\cite{Bouwmeester1997}, entanglement swapping~\cite{Pan1998, Merkouche2022}, multipartite entanglement generation~\cite{Zhong2018}, and quantum repeaters~\cite{Azuma2015}. \bl{In such applications, it is crucial that interfering photons from different pairs be indistinguishable in all auxiliary DoFs, which in path or polarization encoding requires sources that are spectrally pure or ``unentangled'' over the measurement bandwidth. Quantitatively, we can define the spectral purity as $\mathcal{P}=1/K=\sum_n\lambda_n^2$, where the Schmidt number $K$ can be computed through the Schmidt decomposition of the biphoton joint spectral amplitude: $\Phi(\omega_1,\omega_2) = \sum_n \sqrt{\lambda_n} \psi_n(\omega_1)\phi_n(\omega_2)$~\cite{Law2000,Grice2001,Law2004}. Spectral purity $\mathcal{P}=1$ indicates complete separability and ensures that a pure single photon can be heralded by detection of the other photon from the pair.}

\bl{Since frequency itself represents the DoF of interest in frequency-bin encoding, the analysis of spectral purity becomes slightly more nuanced. In general, BFCs produce two types of frequency entanglement: \emph{interbin entanglement} (the coherent superposition of distinct signal-idler frequency-bin pairs) and \emph{intrabin entanglement} (entanglement within a single pair)~\cite{Lu2023c}. Since the former is typically desirable in frequency-bin quantum information while the latter is not (for the same multiphoton interference reasons as in other DoFs), the preferred BFC scenario} entails a pump with bandwidth that is larger than an individual frequency bin but smaller than the bin spacing. However, for microring sources,  where the pump resonance linewidth is (typically) equal to that of the signal and idler, the intraring pump bandwidth will not appreciably exceed that of the generated bins, constraining the maximum achievable spectral purity to 93\% per bin pair with Gaussian pump pulses~\cite{Vernon2017}. To address this limitation, novel approaches have been proposed, including engineering the microring coupling~\cite{Vernon2017, Liu2019, Borghi2024, Burridge2023}, designing combined couplings of multiple microrings~\cite{Alexander2024}, and manipulating the pump pulse~\cite{Christensen2018, Burridge2020}.

For example, in Ref.~\cite{Liu2019} two asymmetric Mach-Zehnder interferometers (AMZIs) couple to the ring [Fig.~\ref{fig:purity}(a)] to independently control the quality factors of the pump and signal-idler. Configured to create a $\pi$-phase difference between adjacent resonances, %each AMZI coupled to alternating resonances, with 
the input AMZI couples to the pump resonance and the output AMZI to the signal and idler resonances. Different coupling coefficients achieve a higher quality factor---and thus narrower linewidth---for the signal and idler relative to the pump. A biphoton spectral purity of (95$\pm$1.5)\% was measured from an unheralded second-order autocorrelation [$g^{(2)}$] measurement~\cite{Faruque2019}, while the estimate obtained from the joint spectral intensity (JSI) [Fig.~\ref{fig:purity}(a)] using stimulated emission tomography (SET)~\cite{Liscidini2013} reached 99.1\%. %, exceeding the 93\% limit for microrings without interferometric coupling~\cite{Vernon2017}.

Another design features an auxiliary ring that enables resonant interferometric coupling to the main ring~\cite{Borghi2024}, %The effective coupling of the main resonator to the bus waveguide is 
influenced by the optical path imbalance of the interferometer arms and the transmission characteristics of the auxiliary ring. An estimated 98.67\% purity has been extracted from an experimentally measured joint temporal intensity (JTI) with this design.

In Fig.~\ref{fig:purity}(b), a primary ring responsible for photon generation is coupled to an auxiliary ring that resonates exclusively at the pump wavelength of the primary ring~\cite{Burridge2023}. An AMZI coupled to the auxiliary ring controls its resonance linewidth, thereby influencing the pump lineshape in the primary resonator. This dual-ring structure enhances the selectivity of pump resonance control, optimizing both heralding efficiency and spectral purity, with the latter estimated at 97\% and 99.1\% from unheralded $g^{(2)}$ and SET-based JSI measurements, respectively. 

A recent design proposes a cascade of microrings that are serially coupled to a bus waveguide with slightly offset resonant frequencies~\cite{Alexander2024}. This arrangement creates an array of closely spaced, overlapping resonance lineshapes across the pump, signal, and idler frequencies. When the pump pulse spectrum is confined within the total bandwidth of the overlapping resonances, the signal and idler become unentangled. A 24-resonator implementation yielded an estimated purity of 99.7\%, based on the JSI measured using SET. %A 24-resonator implementation demonstrated an upper-bound purity of 99.7\% from the SET-based JSI measurement. 
The design is tolerant to variations in global resonance wavelengths caused by fabrication imperfections, with minimal impact on the spectral indistinguishability of photons generated from different devices.

Finally, a scheme based on pump pulse engineering~\cite{Christensen2018}  has also experimentally achieved spectral purity enhancement beyond the 93\% bound without requiring coupling modifications [Fig.~\ref{fig:purity}(c)]~\cite{Burridge2020}. By pumping the ring with two phase-shifted, temporally displaced Gaussians, the resonance lineshape is essentially precompensated, resulting in a flat and broader pump spectrum within the cavity. This approach has reported a purity of 98\% estimated via SET.

\section{State manipulation} 
\label{sec:manipulation}
\subsection{Quantum frequency processor}
Manipulating photons in frequency-bin encoding involves applying controlled phase shifts within and controlled cross-couplings between frequency bins. In the QFP architecture [Fig.~\ref{fig:manipulation}(a)], each pulse shaper applies a precise spectral phase mask to individual frequency-bin modes, while the EOMs act as mode mixers, generating spectral interference across frequency bins~\cite{Lukens2017,Lu2019c,Lu2023a}. While QFPs based on bulk optics have been widely demonstrated, a fully integrated QFP remains unrealized. Nevertheless, integration holds significant promise, offering the potential for lower loss and enhanced scalability. In this subsection, we will explore recent demonstrations of integrated pulse shapers and EOMs with an eye toward uniting these components in a fully integrated QFP.

\subsubsection{Pulse shapers}
Fourier-transform pulse shapers allow for controllable amplitude and phase modulation of the frequency components of an optical signal, typically achieved by spectrally dispersing the signal's spectrum, routing spectral regions to unique amplitude and phase modulators, and then recombining them into a single spatial mode. In the bulk, spectral dispersers are often realized with some combination of diffraction gratings, prisms, and virtually imaged phase arrays, while amplitude and phase modulation is applied by programmable liquid crystal on Si elements~\cite{Weiner2011}. 

\begin{figure}[tb!]
\centering\includegraphics[trim= 0 425 250 0,clip,width=3.2in]{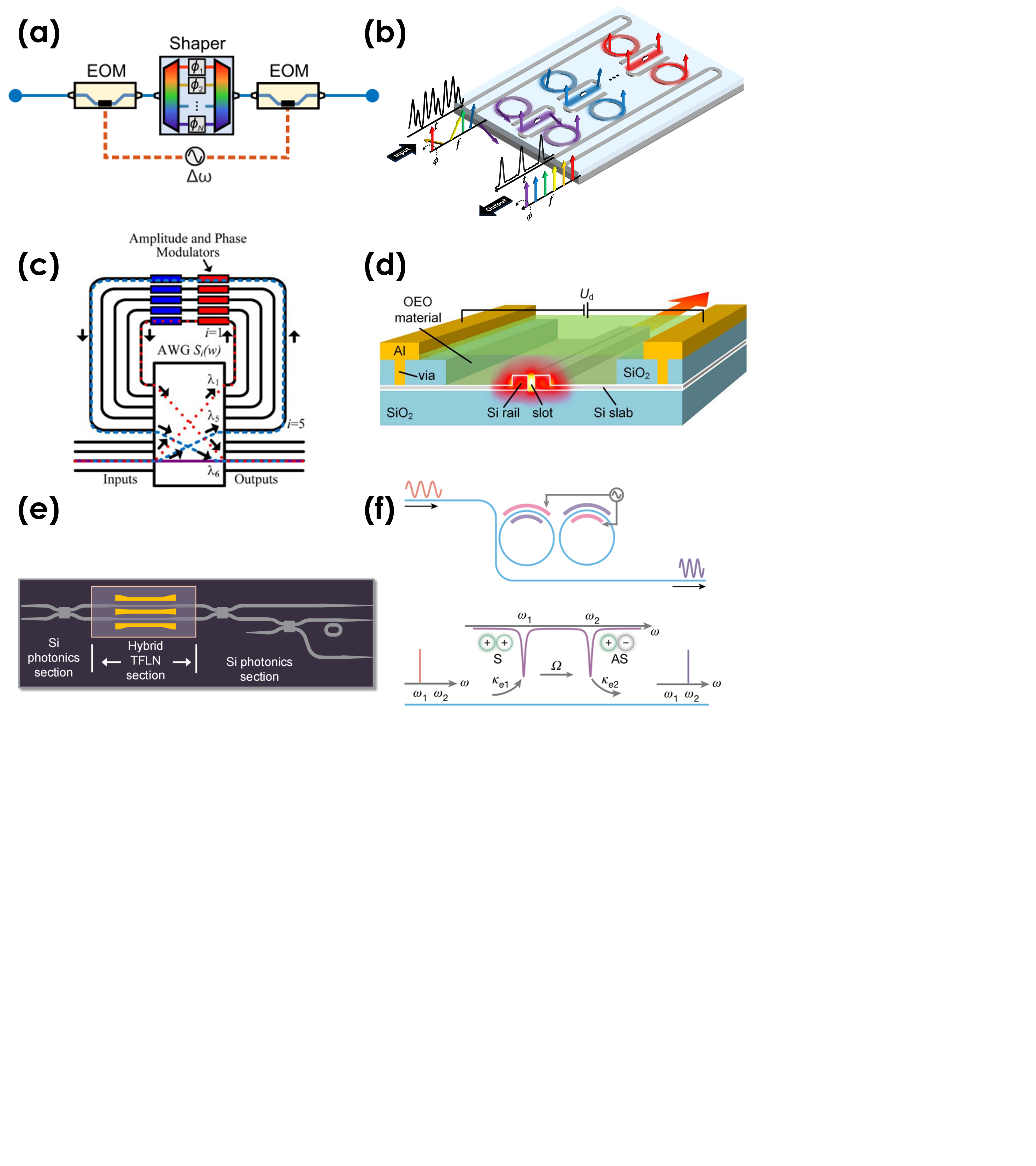}
\caption{Integrated approaches for frequency-bin manipulation. (a)~Three-element QFP~\cite{Lu2023c}. (b)~Six-channel Si microring pulse shaper~\cite{Cohen2024b}. (c)~64-channel SiO$_2$ AWG pulse shaper~\cite{fontaine2008compact}. (d) Hybrid Si-organic EOM~\cite{Kieninger2018}. (e)~Hybrid Si-TFLN modulator for pump control~\cite{Wang2023b}. (f)~TFLN frequency beamsplitter~\cite{Hu2021}. Images reproduced with permission from an
Optica Open Access Publishing Agreement~\cite{Lu2023c, Kieninger2018, Wang2023b}, a
Creative Commons Attribution-NonCommercial-NoDerivatives 4.0 International License
(\url{https://creativecommons.org/licenses/by-nc-nd/4.0/})~\cite{Cohen2024b}, Optica Publishing Group~\cite{fontaine2008compact}, and Springer Nature~\cite{Hu2021}. }
\label{fig:manipulation}
\end{figure}

Integrated spectral shapers follow the same concepts, however with designs that accommodate the intricacies of guided-wave optical physics. For example, integrated spectral dispersers commonly rely on arrayed waveguide grating (AWG) or microring resonator technology. In an AWG, an input field at the object plane enters a free propagation slab and is evenly distributed into an array of waveguides with an incremental optical path length difference, producing a spatial phase gradient. Upon recombining the light from each waveguide in another free propagation slab, the gradient results in spectral focusing at different points in the image plane. By placing output waveguides at appropriate points in the image plane, the AWG functions as a spectral disperser~\cite{smit1996phasar}. In contrast, microring pulse shapers rely on coupling a series of rings to an input bus waveguide, where each ring downloads a specific frequency that can subsequently be phase- or amplitude-tuned before uploading back to an output waveguide, as depicted in Fig.~\ref{fig:manipulation}(b)~\cite{Cohen2024b}. 

Integrated shapers with AWG spectral dispersers have been realized in silica (SiO$_2$)~\cite{fontaine2008compact}, indium phosphide (InP)~\cite{soares2011monolithic, Metcalf2016}, and heterogeneous platforms~\cite{feng2017rapidly}. AWG-based shapers can enable control over a large number of spectral modes; for example, Fig.~\ref{fig:manipulation}(c) illustrates a 64-channel pulse shaper from one study~\cite{fontaine2008compact}, while another has demonstrated up to 100 modes~\cite{soares2011monolithic}. However, these devices are typically limited in spectral resolution to $\gtrsim$10~GHz and loss to $\gtrsim$10~dB, and techniques to improve the resolution result in increased optical loss and phase errors which must be actively compensated. 
Heterogeneous integration can benefit AWG-based shapers by utilizing appropriate material platforms for both the AWGs and amplitude/phase modulators, e.g.,  %for example in Ref.~\cite{feng2017rapidly}, the authors use 
low-loss SiN for the AWG and InP for reconfigurable phase modulation up to $500$~MHz and beyond~\cite{feng2017rapidly}.

Shapers with microring-based dispersers~\cite{Agarwal2006,Khan2010,Wang2015b,Cohen2024b} have proven less common than AWG-based versions, but present some interesting advantages. The microrings are easily tunable with thermo- or electro-optic effects, their geometry is such that spectral resolution improves as loss goes down, and they can achieve low insertion loss on the order of a few dB. %The benefits of reduced loss and footprint are obvious, but 
Importantly, finer spectral resolution allows for an increase in the number of frequency modes within the same optical bandwidth as well as increases the complexity of modulating waveforms available to mix them in the QFP paradigm~\cite{Lu2022a}. Recently, line-by-line pulse shaping has been demonstrated on six lines of a classical electro-optic comb with a spacing of $3$~GHz~\cite{Cohen2024b}. The device contains two sets of multimode microring filter banks with inline phase shifters in between for a total of six individually addressable channels, each with %Microrings are designed using multimode waveguides, giving each channel 
a $3$~dB linewidth of $\sim$900~MHz. 

This same device recently shaped the temporal profile of entangled photons as well~\cite{wu2024chip}. By carefully selecting mode sets to carve the biphoton spectrum, spectral phase control could be realized in a $3 \times 3$ or $6 \times 6$ Hilbert space of frequency-bin entangled qudits at a bin separation of $3$~GHz. At such fine bin separations, the pulse shaper can control relatively long temporal features (on the order of $\sim$1~ns), allowing for direct measurement of the temporal correlation function with commercial single-photon detectors. %This work represents a significant milestone in the field, showcasing the first successful demonstration of pulse shaping for quantum light using an integrated shaper. 

\subsubsection{Electro-optic modulators}
Single-tone optical phase modulation scatters the carrier light into sidebands at harmonics of the modulation frequency, as described by the Jacobi--Anger expansion. %, with amplitudes related to the modulation index. EOMs are therefore critical elements in mixing frequency-bin modes. 
Characteristics of an ideal EOM are low insertion loss, low chirp, large electro-optic bandwidth, a compact footprint, and a low half-wave voltage $V_\pi$\bl{---the voltage required to induce a $\pi$ phase shift on the optical field.} Here we focus primarily on Pockels effect modulation (in contrast to common integrated modulation methods like plasma dispersion~\cite{Reed2010, Rahim2021, Sinatkas2021}) because it produces index changes that are linearly proportional to the applied electric field and imparts no intrinsic optical loss. %, does not intrinsically consume electrical power, and shows particular promise for frequency-bin photonics.
Notable integrated electro-optic materials in this category are lithium niobate (LiNbO$_3$)~\cite{Boes2023}, barium titanate (BaTiO$_3$)~\cite{Karvounis2020}, galium arsenide (GaAs) and AlGaAs \cite{Baboux2023}, and organic electro-optic polymers [Fig.~\ref{fig:manipulation}(d)]~\cite{Kieninger2018}. 

LiNbO$_3$ modulators have long been used in optical networks and telecommunications applications~\cite{Wooten2000}. Waveguides are formed in bulk via in-diffusion of ions and are weakly guiding, resulting in bulky devices with limited $V_{\pi}$ and operational speed. Processing TFLN, with thicknesses at hundreds of nm, has been extensively researched to overcome these limitations~\cite{Boes2023}, with TFLN modulators already successfully employed in frequency-bin quantum manipulation. For instance, Ref.~\cite{Zhu2022} uses a double-pass TFLN EOM (with $V_{\pi} \approx$ 2.3--2.8~V at modulation frequencies of 10--40 GHz) to achieve spectral shearing of heralded single photons, reaching a record-high 641~GHz frequency shift. By pairing the EOM with a dispersion module, spectral compression of single photons is also achieved. Both techniques show promise for addressing frequency and bandwidth mismatches between disparate quantum systems.

Unetched TFLN bonded to Si has been used to integrate TFLN modulators with high-Q resonators to control biphoton generation [Fig.~\ref{fig:manipulation}(e)]~\cite{Wang2023b}. %Here, the authors use silicon photonics to route pump light through an MZI and high-Q resonator. 
When TFLN is bonded above the Si waveguide, the optical mode is hybridized, achieving a high mode fraction in the TFLN (up to 80\%) while preserving guidance in the underlying Si layer. In this way, the authors realize a TFLN intensity modulator with $1.8$~dB insertion loss, \bl{voltage-length product} $V_{\pi}L \approx 3.5$~V~cm, $>$100~GHz $3$~dB bandwidth, and $\sim$26~dB extinction, which they use to carve continuous-wave pump light into pulses of various duration to initiate SFWM in a microring. By reducing the pump pulse duration, the authors produce biphotons with a purity reaching the 93\% bound as measured by SET. 

Binary and ternary III--V semiconductor materials like GaAs and AlGaAs are also promising for their intrinsic linear electro-optic effect. Although the electro-optic coefficient is generally $\sim$10 times lower than that of TFLN, III--Vs offer a direct bandgap facilitating lasers and photodetectors, enjoy a large index contrast for high-density integration, and offer a broader transparency window than Si. Additionally, polymers with active organic molecules %possessing large hyperpolarizability 
have shown promising results after deposition on hybrid Si photonics platforms. The high polarizability of these polymers leads to strong electro-optic coefficients $r_{33}>$300~pm/V~\cite{Kieninger2018}. Furthermore, hybrid organic polymer modulators have been shown to accommodate speeds beyond $100$~GHz~\cite{lee2002broadband, Alloatti2014, Ummethala2021}.

%\rd{On another front, acousto-optic modulators are emerging as promising alternatives to EOMs for on-chip phase modulation. TALK ABOUT PAPERS HERE. What would the main distinctions be compared to EOMs?}
\bl{On another front, acousto-optic modulators (AOMs) are emerging as promising alternatives to EOMs for on-chip phase modulation. The acousto-optic effect, particularly through Brillouin scattering, involves altering the refractive index of an optical material due to the strain induced by acoustic waves, enabling the manipulation of light as it travels through the medium. Devices based on this phenomenon are well developed in bulk-crystal AOMs~\cite{savage2010acousto}. Integrated AOMs utilize $\upmu$m-scale waveguides to achieve efficient acousto-optical coupling between GHz-frequency acoustic waves and optical signals. These surface acoustic waves are generated by electrically driven interdigital transducers placed on piezoelectric materials such as LiNbO$_3$~\cite{shao2020integrated} or AlN~\cite{zhao2022enabling,Zhou2024}. The waveguides are typically suspended to minimize acoustic losses to the substrate and enhance the acousto-optic interaction.}

\bl{Notably, AOMs fabricated on the AlN-on-SOI platform, designed to incorporate coexisting acoustic and optical resonances, have demonstrated impressive performance with $V_{\pi} \approx 19$~mV, operating within the optical bandwidths defined by the nanophotonic cavity resonance linewidth~\cite{zhao2022enabling}. Additionally, AOMs featuring linear waveguide segments~\cite{Zhou2024} that support broad optical bandwidths exceeding 100~nm have been developed with $V_{\pi}L \approx 0.1$~V cm, enabling applications in microwave-to-optical transduction. However, typical AOMs require acoustic resonance to facilitate strong acousto-optic coupling and therefore suffer limited RF bandwidths, although there are paths to overcome this via multiple or chirped interdigital transducers~\cite{fall2017generation}.}
% Although the modulation efficiency currently lags  optically resonant AOMs, it can be enhanced by cascading multiple linear AOM segments, thanks to their exceptionally low optical losses.}

\subsection{Non-QFP approaches}
In addition to the pulse shapers and single-pass EOMs comprising the QFP paradigm, non-QFP techniques for frequency-bin control are also being explored. In this subsection, we focus on two classes of alternatives demonstrated so far in integrated photonics: (i)~the electro-optic photonic molecule, which relies on coupled cavities to reduce the number of interacting modes compared to a traditional EOM; and (ii)~integrated $\chi^{(2)}$ and $\chi^{(3)}$ parametric processes mediated by classical optical pump fields.

\subsubsection{Photonic molecule}
Coupled resonant cavities enable coherent, dynamic control of the supported energy levels via fast phase modulation within each cavity~~\cite{Zhang2019b, Buddhiraju2021, Hu2021}. %In this approach,  induces coherent coupling between the photonic modes. 
By precisely controlling the frequency and amplitude of the radio-frequency (RF) driving waveform, unitary frequency-bin transformations can be achieved, provided the photon lifetime is longer than the time needed to drive the system from one state to the other. In this context, TFLN is an excellent contender as it supports low waveguide propagation losses \cite{shams2022reduced} and ultrafast modulation speeds theoretically reaching fs timescales \cite{zhang2022systematic}.

Indeed, Ref.~\cite{Zhang2019b} realizes a TFLN photonic molecule with a system of two coupled microrings emulating a two-level atomic system, achieving %. The low loss and efficient modulation of the TFLN platform allow them to realize a device with 
large electrical bandwidth ($>$30~GHz) %, high modulation efficiency ($0.5$~GHz/V), 
and long photon lifetime ($\sim$2~ns). Using dynamic pulsed modulation, precise control over coupling between the resonant modes of the photonic molecule is obtained,  enabling phenomena like Rabi oscillations and Ramsey interference. 

Reference~\cite{Hu2021} extends the photonic molecule concept for %show a two-resonator photonic molecule, as shown in Fig.~\ref{fig:manipulation}(f), is capable of 
bidirectional frequency shifting with $\sim$90\% conversion efficiency and $\sim$30~dB carrier suppression under monotone RF modulation [Fig.~\ref{fig:manipulation}(f)]. %Furthermore, the applied RF power can control the shifting ratio allowing them to realize frequency beam splitters ($50/50$ splitting) and anything in between.Unlike pure phase modulation methods, the resonant nature of the photonic molecule causes the device to operate only on a select pair of modes. 
Furthermore, the device can be configured to function as a frequency beamsplitter (achieving any desired splitting ratio by adjusting the applied RF power). The resonant nature of the photonic molecule confines mixing to a specific pair of modes, preventing the creation of unwanted frequency sidebands associated with single-pass EOMs.
Moreover, the authors present a cascaded frequency shifter with three microrings to enable frequency shifts of $120$~GHz with only 30~GHz RF excitation. These experiments and concepts could pave the way for high-dimensional frequency-bin transformations with photonic molecules. 

\subsubsection{Nonlinear optics}
Nonlinear frequency mixing is one of the earliest and most straightforward methods for manipulating the frequency components of single photons. Originally established for quantum frequency conversion~\cite{Kumar1990, Huang1992}, this approach enables shifting a photon’s frequency while preserving its quantum information. When the information is directly encoded in the frequency DoF, these frequency shifts \bl{function as unitary transformations between distinct spectral basis states, with weights controllable by the interaction strength~\cite{Clemmen2016}.} %allow for the coherent transfer of population between different frequency modes. This transfer can be either partial or complete, depending on the interaction strength, and function as unitary transformations between distinct spectral basis states~\cite{Clemmen2016}.} 
These transformations can be realized through single-pump $\chi^{(2)}$ processes like sum-frequency generation (SFG) and difference-frequency generation (DFG) or dual-pump $\chi^{(3)}$ processes such as Bragg scattering four-wave mixing (BS-FWM).

Most experimental demonstrations of nonlinear frequency mixing for frequency-encoded photons have focused on simple two-dimensional operations, such as frequency beamsplitters~\cite{Kobayashi2016, Kobayashi2017, Clemmen2016} and their use in Hong--Ou--Mandel interference between spectrally distinct photons~\cite{Kobayashi2016, Joshi2020}. However, recent efforts have begun to extend these principles to higher dimensions, %with theoretical and experimental advances proposing 
such as generalization of BS-FWM processes to $N$-way frequency beamsplitters~\cite{Oliver2023}. Furthermore, a growing convergence between temporal-mode and frequency-bin encoding has highlighted the versatility of nonlinear mixing. By exploiting spectrally shaped pump fields and engineered poling structures, multi-output quantum pulse gates~\cite{Serino2023}, originally developed for temporal modes, have shown potential for manipulating high-dimensional frequency-bin states~\cite{Serino2024}.

Historically, these frequency mixing demonstrations have primarily utilized bulk nonlinear waveguides and fibers. Periodically poled lithium niobate (PPLN) waveguides, for example, are frequently employed in SFG and DFG experiments, while BS-FWM processes typically leverage nonlinear photonic-crystal fibers. However, recent advances in nanophotonic platforms are poised to transform this field. Single-photon frequency upconversion via SFG has now been achieved using both TFLN waveguides~\cite{Wang2023a} and microrings~\cite{Chen2021}, while SiN microrings are increasingly explored for on-chip BS-FWM~\cite{Li2016, Singh2019}. Despite this progress, fiber-to-chip coupling losses remain a primary challenge, and engineering phase-matching conditions is essential to boosting internal conversion efficiencies, motivating continued innovation to realize on-chip frequency-domain quantum circuits using nonlinear optics.

%- If there is any good paper that Nils can help point out, I wonder if we could also mention AOM for frequency mixing. Linbo/Loncar's TFLN paper at least have shown frequency shifting with very good contrast on-chip.

\section{Hyperentanglement}
\label{sec:hyperent}
Hyperentanglement describes quantum particles that are simultaneously entangled across multiple \emph{independent} DoFs and has shown promise in various quantum communication tasks~\cite{Simon2002, Barreiro2008, Graham2015}. Given the ubiquity of frequency entanglement in most photon-pair generation processes, it is no surprise that hyperentanglement involving frequency and an additional DoF has been widely investigated. %In fact, for those who solely focus on entanglement in other DoFs, pulse pumping is often used to eliminate unwanted spectral correlation across the whole spectrum~\cite{Grice1998, Grice2001}, especially in applications like multiphoton interference~\cite{Pan2012}. 
In the following subsections, we will highlight three different types of such hyperentanglement, with a particular emphasis on those demonstrated in integrated photonics. %These platforms provide scalable and efficient methods for leveraging hyperentangled states in advanced quantum technologies.

\begin{figure}[b!]
    \centering
    \includegraphics[width=\columnwidth]{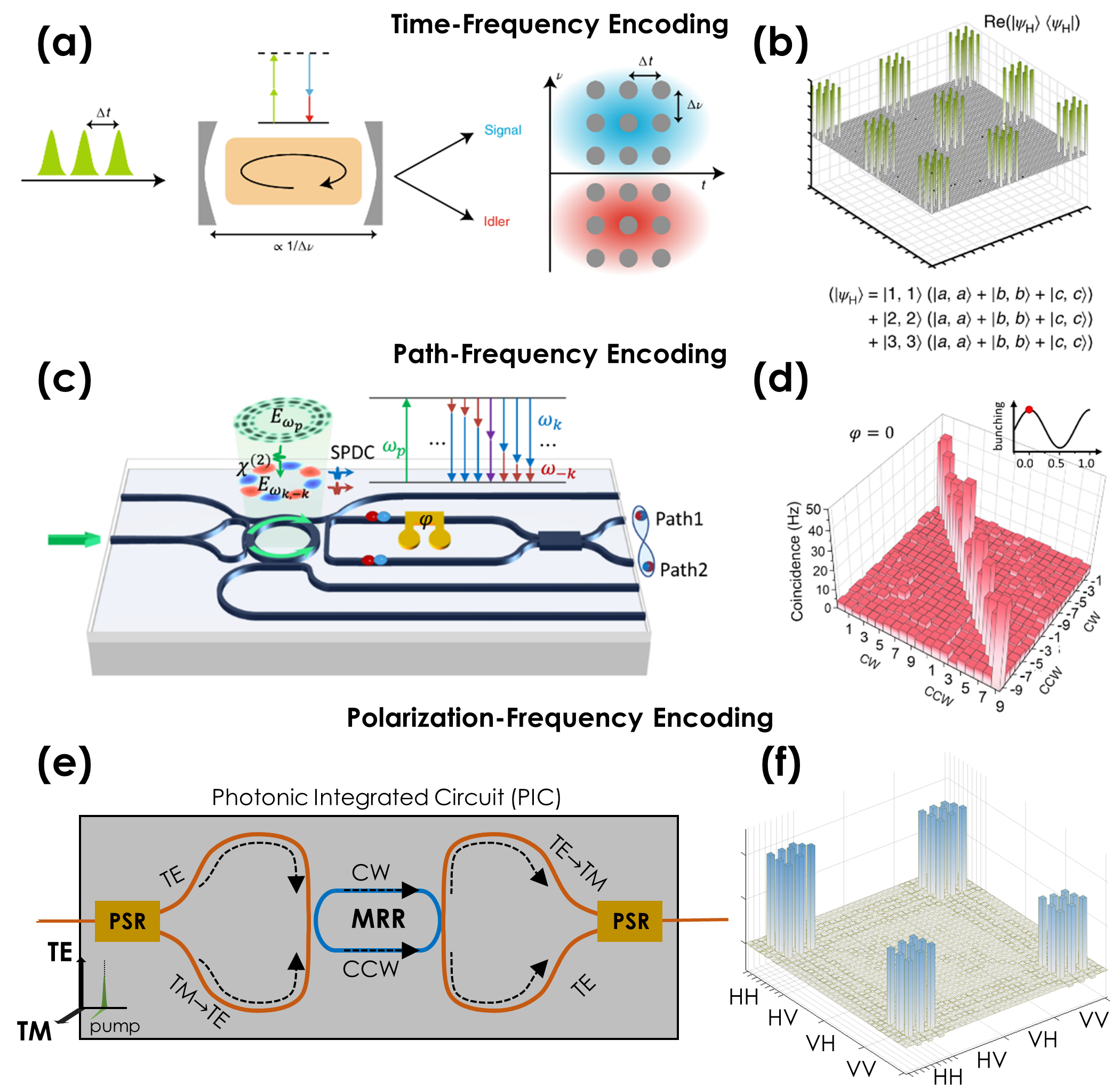}
    \caption{Examples of hyperentanglement involving frequency and other DoFs. (a) Time-frequency encoding scheme and (b) corresponding representative density matrix~\cite{Reimer2019}. (c) Path-frequency encoding scheme with (d) associated correlation matrix~\cite{Zhang2023}. (e) Polarization-frequency encoding scheme~\cite{Miloshevsky2024} and (f) example density matrix from a separate study~\cite{Lu2023b}. Images reproduced with permission from an Optica Open Access Publishing Agreement~\cite{Miloshevsky2024}, a Creative Commons Attribution-NonCommercial-NoDerivatives 4.0 International License (\url{https://creativecommons.org/licenses/by-nc-nd/4.0/})~\cite{Zhang2023}, Optica Publishing Group~\cite{Lu2023b}, and Springer Nature~\cite{Reimer2019}.}
    \label{fig:Hyperentanglement}
\end{figure}

%NOTE: Mention they are naturally hyperentnagled, but not all of them have verified that.

\subsection{Time-frequency encoding}
%The first set of examples involve time-frequency hyperentanglement. 
The combination of time and frequency may seem counterintuitive at first glance, since these DoFs are Fourier conjugates and thus intrinsically constrained by the uncertainty principle. However, one can operate in a regime far from this limit, for example by exciting a nonlinear medium with pump pulses spaced such that their temporal separation (i.e., the time-bin spacing) is much greater than the inverse frequency-bin linewidth [Fig.~\ref{fig:Hyperentanglement}(a)]~\cite{Reimer2019}. This strategy effectively decouples the two DoFs, enabling independent manipulation of both time-bin and frequency-bin entanglement.

In two microring-based examples (Hydex~\cite{Reimer2019} and SiN~\cite{Imany2019}), modulation of the pump controls the time-bin structure, while the natural resonance characteristics of the microring define the frequency bins. Figure~\ref{fig:Hyperentanglement}(b) presents a representative density matrix, where photon pairs exhibit simultaneous entanglement across three frequency-bin and three time-bin pairs~\cite{Reimer2019}. Given the theoretically unbounded number of modes accessible in both DoFs, time-frequency hyperentanglement offers a straightforward route to expanding into larger Hilbert space dimensions. For instance, manipulation of time-frequency hyperentangled states has enabled synthesis of a two-photon, 32-dimensional GHZ state~\cite{Imany2019}.

While the \emph{generation} of such high-dimensional quantum states appears relatively simple for time-frequency encoding, the \emph{manipulation} of these states presents greater challenges. Both examples above demonstrate specific, deterministic unitaries---namely, quantum-controlled gates between two DoFs within a single photon---using techniques like dispersion-induced frequency-to-time mapping and high-speed electro-optic modulation. However, the range and complexity of such controlled operations remain limited and are not yet compatible with fully integrated systems. %This constraint is not unique to time-frequency encoding but is a common challenge across all photonic hyperencoding schemes.

\subsection{Path-frequency encoding}
\label{pathfreq}
Perhaps the most obvious approach for path-frequency hyperentanglement involves coherently pumping multiple microrings located in different optical paths. %[in contrast to, e.g., Fig.~\ref{fig:MRR}(d) where the source microrings couple to the same bus waveguide]. %With sufficiently low pump power, at most a single pair of photons is emitted across all paths with high probability. 
The inherent frequency entanglement within each microring is augmented by path entanglement arising from the use of multiple bus waveguides. %However, fabrication tolerances---particularly variations in FSR---make it challenging to ensure similar characteristics across multiple microrings.
For two-dimensional path entanglement, a simpler alternative is to %This issue can be effectively addressed by bidirectionally 
pump a single microring bidirectionally, as shown in the AlN design of Fig.~\ref{fig:Hyperentanglement}(c)~\cite{Zhang2023}. In this configuration,
photon pairs generated by the clockwise (CW) pump propagate into the lower waveguide, while those generated by the counterclockwise (CCW) pump propagate into the upper waveguide. %photon pairs are emitted into the upper waveguide from the pump in one propagation direction and into the lower waveguide from the pump in the counter-propagating direction, creating two-dimensional ($d=2$) path entanglement. %However, this approach restricts the path entanglement to two dimensions. 

Heaters and Mach--Zehnder interferometers allow for two photons to travel in either the same or different paths. When the photon pairs travel along the same path, their correlations in both frequency and path DoFs are depicted in Fig.~\ref{fig:Hyperentanglement}(d).
Although not explicitly confirmed in this study, previous demonstrations with similar microring structures suggest that frequency entanglement is inherently present, positioning this source as a natural candidate for future path-frequency applications. To extend the dimensionality in the path domain beyond $d=2$, multiple microrings in distinct optical paths would ultimately be required.%, and ensuring identical spectral distributions across these microrings is essential for achieving high-quality hyperentangled states.

\subsection{Polarization-frequency encoding}
While frequency entanglement can be easily achieved with microrings, on-chip polarization entanglement presents significant challenges. Integrated waveguides are typically optimized for a single polarization, as achieving equivalent performance for both transverse-electric and transverse-magnetic polarizations is difficult. Previous efforts in this area have used Hydex microrings supporting orthogonal polarization resonances~\cite{reimer2015}, AlGaAs waveguides accommodating both polarizations~\cite{horn2013,kultavewuti2017}, and setups involving the polarization-rotation-based combination of photon pairs from two similar Si waveguides, operating either in series~\cite{matsuda2012} or parallel~\cite{olislager2013}.

In discrete fiber optics, a particularly effective approach for polarization-frequency hyperentanglement has proven to be bidirectional pumping in a Sagnac configuration with polarization-rotating elements%, with several demonstrations . This technique has been used to pump nonlinear sources bidirectionally in a fiber-based Sagnac loop, where the polarization of photon pairs in one direction is rotated to generate polarization entanglement
~\cite{fan2007,vergyris2017,Alshowkan2022}. %These methods resemble earlier demonstrations of path-frequency entanglement in Sagnac loops, with the key difference being the replacement of the 50:50 beamsplitter at the input with a polarizing beamsplitter and 90$^\circ$ rotation in one of the arms, effectively converting path entanglement into polarization entanglement. 
This method provides the distinct benefit of achieving similar characteristics for both output polarizations.
Similarly, microrings placed within a fiber Sagnac loop have been employed to realize broadband polarization entanglement spanning multiple energy-matched resonances, facilitating hyperentanglement in both polarization and frequency bins~\cite{suo2015,wen2023}. However, these efforts have relied on fiber-based Sagnac loops and were not fully integrated.

A recent bidirectionally pumped Si microring with on-chip polarization splitter-rotators\bl{---a component that separates orthogonal polarizations and rotates one by 90$^\circ$~\cite{dai2013,pruessner2024}---}marks the first integrated demonstration of this approach [Fig.~\ref{fig:Hyperentanglement}(e)]~\cite{Miloshevsky2024}. The authors perform full tomography of the polarization state for 117 pairs of signal and idler resonances with a frequency spacing of 38.4~GHz. Although hyperentanglement has not been explicitly verified, the intrinsic frequency entanglement is expected to coexist with the observed polarization entanglement. Figure~\ref{fig:Hyperentanglement}(f) shows a 36-dimensional polarization-frequency hyperentangled state from a PPLN source in a fiber-based loop~\cite{Alshowkan2022}, similar to what we would expect from the integrated source in Fig.~\ref{fig:Hyperentanglement}(e) after full characterization.

\section{Conclusion}
\label{sec:conclusion}
As the examples featured in this Perspective have highlighted, the last five years have observed a fundamental transformation in integrated frequency-bin photonics. What began for some of us as a grand vision for quantum information processing~\cite{Lukens2017}---whose sole experimentally validated on-chip components were, at that time, single-ring sources~\cite{Clemmen2009, Azzini2012, Grassani2015, Ramelow2015, Reimer2016}---has since exploded with encouraging on-chip developments: multiring sources for controllable state synthesis and interferometric techniques for enhanced spectral purity (Sec.~\ref{sec:generation}); integrated quantum pulse shapers and on-chip modulators (Sec.~\ref{sec:manipulation}); and hyperentangled sources merging frequency bins with time, path, and polarization (Sec.~\ref{sec:hyperent}). Perhaps the most striking commonality weaving through these advances is the microring. Visually represented in all panels of Figs.~\ref{fig:MRR} and \ref{fig:purity} and two panels each in Figs.~\ref{fig:manipulation} and \ref{fig:Hyperentanglement}, the natural synergy between frequency-bin encoding and microring resonators, posited in Sec.~\ref{sec:intro}, is confirmed by example.

\bl{The resonant nature of microring circuits do make them particularly sensitive to fabrication variations, which can significantly impact their performance. For instance, deviations in the quality factors from their ideal values can substantially lower the pair generation rate, while linewidth variations across different microrings can degrade the quality of synthesized frequency-bin states in multiring sources (cf. Secs.~\ref{multipleMRR} and \ref{pathfreq}). However, the scalability of photonic integration does mitigate this variation to some extent by enabling the fabrication of many circuit copies and designs~~\cite{lu2017performance,bogaerts2019layout}, allowing researchers to test and handpick optimal devices. Additionally, strategies that leverage statistical variations with many devices (like the 24-ring cascade of Ref.~\cite{Alexander2024}) or multiple uses of a single device [like bidirectional pumping of a single ring in Fig.~\ref{fig:Hyperentanglement}(c,e)] have revealed pathways to reduce fabrication dependencies through circuit engineering.}
%Additionally, post-fabrication tuning techniques,} \rd{such as XXX and YYY}, \bl{can be employed to fine-tune coupling conditions and mitigate fabrication-induced discrepancies. These strategies collectively help maintain the performance and quality of frequency-bin states despite fabrication challenges. }

Yet it is perhaps what microrings alone \emph{cannot} accomplish that will define the future of frequency-bin photonics. For while microrings can be realized in virtually any integrated platform, efficient electro-optic modulation is possible only in specific materials. As discussed in Sec.~\ref{sec:manipulation}, the highest-performing modulators applied to quantum light have to date have been realized on TFLN---unsurprising in light of bulk LiNbO$_3$'s longstanding dominance in the discrete-EOM market. 
\bl{However, TFLN is not compatible with CMOS foundries, the ecosystem in which the most advanced on-chip microring pulse shapers (in terms of controllability and channel number) have been realized~\cite{Wang2015b,Cohen2024b}. In our experience, the mismatch between the best materials for a complex pulse shaper and the best materials for an ideal EOM has proven the most challenging impediment to a fully on-chip QFP. Nevertheless, the sharp distinction between CMOS and non-CMOS materials is softening, as research in heterogeneous integration, where systems can leverage the strengths of multiple material platforms, has begun to thrive~\cite{elshaari2020hybrid,shekhar2024roadmapping}.}

Consequently, the pursuit of complete frequency-bin processing circuits on chip faces a nontrivial decision. Should we pursue CMOS solutions, incorporating compatible materials as needed for modulation? Or should we focus on the most promising materials for monolithic performance and devote efforts toward scalable manufacturing? Or could the best option lie in hybrid integration, with devices on multiple platforms combined through, e.g., wafer bonding or photonic wirebonds? Each of these paths faces its own tradeoffs, with no obvious winner in our view.

Finally, although we have focused on integrated photonics for coherent frequency-bin processing---i.e., as qubits and qudits---we recognize the historical and well-deserved leadership of path-encoded qubits for on-chip quantum photonics. Given the ability of frequency bins to blend smoothly with other on-chip DoFs (Sec.~\ref{sec:hyperent}), it is likely that the capabilities summarized here can play important roles even for qubits encoded in other DoFs, where the goal might be to utilize frequency differences (as in multiplexing~\cite{Miloshevsky2024}) or to erase them (as in photon interference~\cite{Zhu2022}). Accordingly, the future of frequency bins appears extremely promising in on-chip quantum photonics---in whatever assignment they are asked to complete.

\begin{acknowledgments}
We thank A. Miloshevsky and K. Wu for valuable discussions. A portion of this work was performed at Oak Ridge National Laboratory, operated by UT-Battelle for the U.S. Department of Energy under contract no. DE-AC05-00OR22725. The Quantum Collaborative, led by Arizona State University, provided valuable expertise and resources for this project. Sandia National Laboratories is a multi-mission laboratory managed and operated by National Technology \& Engineering Solutions of Sandia, LLC (NTESS), a wholly owned subsidiary of Honeywell International Inc., for the U.S. Department of Energy’s National Nuclear Security Administration (DOE/NNSA) under contract DE-NA0003525. This written work is authored by an employee of NTESS. The employee, not NTESS, owns the right, title and interest in and to the written work and is responsible for its contents. Any subjective views or opinions that might be expressed in the written work do not necessarily represent the views of the U.S. Government. The publisher acknowledges that the U.S. Government retains a non-exclusive, paid-up, irrevocable, world-wide license to publish or reproduce the published form of this written work or allow others to do so, for U.S. Government purposes. The DOE will provide public access to results of federally sponsored research in accordance with the DOE Public Access Plan. This work was done independently of, and is not sponsored by, the U.S. Department of Energy or Sandia National Laboratories. Funding was provided by the U.S. Department of Energy (ERKJ432, DE-SC0024257); Oak Ridge National Laboratory (Laboratory Directed Research and Development Program).
\end{acknowledgments}

%\bibliography{References}


\end{document}